\documentclass[twocolumn,aps,showpacs,prb]{revtex4-1}

\usepackage{amsmath,amssymb,graphics,epsfig,epstopdf,color,multirow,array,verbatim,ulem,braket,tabularx}
\usepackage[colorlinks,linkcolor=blue,citecolor=blue,urlcolor=blue]{hyperref}
\usepackage{color}
\usepackage{graphicx}
\usepackage{epstopdf}
\usepackage{amsmath}
\usepackage{multirow}
\usepackage{ulem}

\begin{document}

\title{Topological properties of Mo$_{2}$C and W$_{2}$C superconductors}

\author{Ning-Ning Zhao$^{1}$}
\author{Peng-Jie Guo$^{2}$}
\author{Xiao-Qin Lu$^{1}$}
\author{Qiang Han$^{1}$}
\author{Kai Liu$^{1}$}\email{kliu@ruc.edu.cn}
\author{Zhong-Yi Lu$^{1}$}\email{zlu@ruc.edu.cn}

\affiliation{$^{1}$Department of Physics and Beijing Key Laboratory of Opto-electronic Functional Materials $\&$ Micro-nano Devices, Renmin University of China, Beijing 100872, China}
\affiliation{$^{2}$Songshan Lake Materials Laboratory, Dongguan, Guangdong 523808, China}

\date{\today}

\begin{abstract}

The topological electronic properties of orthorhombic-phase Mo$_{2}$C and W$_{2}$C superconductors have been studied based on first-principles electronic structure calculations. Our studies show that both Mo$_{2}$C and W$_{2}$C are three-dimensional strong topological insulators defined on curved Fermi levels. The topological surface states on the (001) surface of Mo$_{2}$C right cross the Fermi level, while those of W$_{2}$C pass through the Fermi level with slight electron doping. These surface states hold helical spin textures and can be induced to become superconducting via a proximity effect, giving rise to an equivalent $p+ip$ type superconductivity. Our results show that Mo$_{2}$C and W$_{2}$C can provide a promising platform for exploring topological superconductivity and Majorana zero modes.

\end{abstract}


\maketitle

\section{INTRODUCTION}

The coexistence of superconductivity and topological band structure in a material may induce novel physical phenomena such as topological superconductivity \cite{Ref_1,Ref_2}. Topological superconductors can host Majorana zero modes \cite{Ref_3} that obey non-Abelian statistics and have a potential application in topology-protected quantum computation \cite{Ref_4}. In a conventional $s$-wave superconductor, spins with two degrees are a main obstacle for realizing  Majorana zero modes. In contrast, a spin-less superconductor provides an ideal platform for realizing Majorana zero modes, which can either locate at the edge of a one-dimensional (1D) $p$-wave topological superconductor or bind to the magnetic vortices in a two-dimensional (2D) $p+ip$ topological superconductor \cite{Ref_3}.
Recently, the exploration of topological superconductor has attracted intensive attention in condensed matter physics and materials communities.
Nevertheless, the $p$-wave superconductor viewed as intrinsic topological superconductor is very rare in nature \cite{Ref_5}.

Several approaches for realizing equivalent topological superconductivity have been proposed based on Fu and Kane's work~\cite{Ref_6} and subsequently confirmed by several experiments~\cite{Ref_7Nanowire,Ref_8Featom,Ref_Bi2Se3,Ref_9Bi2Se3/NbSe2,Ref_exFeSeTe1,Ref_exFeSeTe2,Ref_exFeSeTe3,
Ref_LiFeOHFeSe}. The first way is to construct a heterostructure composed of topological insulator and $s$-wave superconductor, in which the superconductivity in helical topological surface states can be induced via a proximity effect at the interface \cite{Ref_6,Ref_9Bi2Se3/NbSe2}. Such a heterostructure, however, is confronted with complex interface effect. The second way is to make a topological material superconducting below a critical temperature by doping or to tune a superconducting material into a topological phase via doping \cite{Ref_10,Ref_11_MFvortice,Ref_CuxBi2Se3,Ref_NbxBi2Se3}. Nevertheless, the inhomogeneity caused by doping is unavoidable and the induced superconducting transition temperature ($T_\text{c}$) is usually very low \cite{Ref_12syn}. In addition, a topological material that can be induced into superconducting phase by pressure or point contact may also realize topological superconductivity \cite{Ref_Sb2Te3_pres,Ref_Cd3As2_SC}. So far, an ideal approach is to find  a single compound that possesses both intrinsic superconductivity and nontrivial topological electronic properties \cite{Ref_exFeSeTe1,Ref_exFeSeTe2,Ref_exFeSeTe3,Ref_FeSeTe_film,Ref_FeSeTecal1,Ref_FeSeTecal2,Ref_LiFeAs}. In such a single compound, the Dirac-type surface states can be guaranteed by topological protection and the superconductivity in the surface states can be induced via a proximity effect due to the intrinsic bulk superconductivity, resulting in an equivalent $p+ip$ type topological superconductivity at the surface.

Following this strategy, we have studied the $\xi$-Fe$_{2}$N-type superconducting carbides Mo$_{2}$C and W$_{2}$C by using first-principles electronic structure calculations. We find that both Mo$_{2}$C and W$_{2}$C are a class of materials with nontrivial topological electronic band structures. Furthermore, previous experiments have shown that the superconducting transition temperatures ($T_\text{c}$) of Mo$_{2}$C and W$_{2}$C are 7.3 K \cite{Ref_Tc_Mo} and 4.05 K \cite{Ref_Tc_W}, respectively.

\section{Method}

The electronic structures of the $\xi$-Fe$_{2}$N-type (orthorhombic-phase) Mo$_{2}$C and W$_{2}$C were studied based on the first-principles electronic structure calculations\cite{Ref_Dft}. The projector augmented wave (PAW) method \cite{Ref_paw} as implemented in the VASP package \cite{Ref_vasp1,Ref_vasp2} was used to describe the core electrons as well as the interaction between the core and the valence electrons. The generalized gradient approximation (GGA)\cite{Ref_gga} of Perdew-Burke-Ernzernof  type was adopted for the exchange correlation functional. The kinetic energy cutoff of the plane-wave basis was set to be 520 eV. A $11\times9\times11$ $k$-point mesh for Brillouin zone (BZ) sampling and the Gaussian smearing method with a width of 0.05 eV for Fermi surface broadening were utilized. Both cell parameters and internal atomic positions were fully relaxed until the forces on all atoms were smaller than 0.01 eV/{\AA}. Once the equilibrium structures were obtained, the electronic structures were further studied with the inclusion of spin-orbit coupling (SOC). The surface states in the projected 2D BZ were studied by using the WannierTools package\cite{Ref_wantool}. The tight-binding Hamiltonian of semi-infinite system was constructed by the maximally localized Wannier functions\cite{Ref_wanfun1,Ref_wanfun2} for the outmost $s$, $p$, and $d$ orbitals of the Mo and W atoms and the outmost $s$ and $p$ orbitals of the C atom generated by the first-principles calculations. The surface states were obtained from the surface Green's function of the semi-infinite system.
To simulate the charge doping effect, the total electron number in the unit cell was changed in self-consistent and band structure calculations.

\begin{figure}[tb]
\includegraphics[angle=0,scale=0.12]{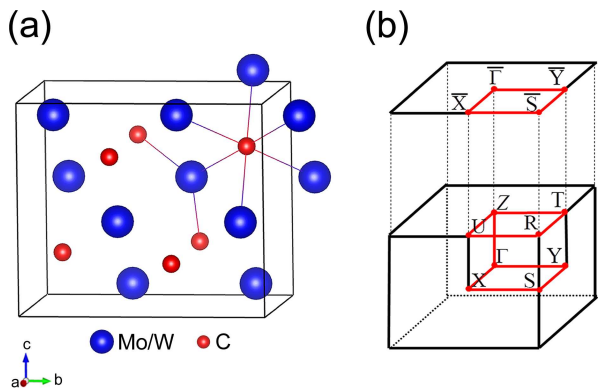}
\caption{(Color online) (a) Crystal structure of orthorhombic-phase Mo$_{2}$C and W$_{2}$C. (b) Bulk Brillouin zone (BZ) and projected two-dimensional BZ of the (001) surface. The red lines indicate the high-symmetry paths in BZ for band structure calculations.}
\label{fig1}
\end{figure}

\section{Results}

The transition metal carbides both Mo$_{2}$C and W$_{2}$C have two phases, namely orthorhombic and hexagonal phases. Here we focus on their orthorhombic phase, which adopts the $\xi$-Fe$_{2}$N crystal structure with space group \textit{Pbcn}. As illustrated in Fig. \ref{fig1}(a), the primitive cell includes eight transition metal atoms and four carbon atoms. Each Mo or W atom possesses three C neighbors, while each C atom is coordinated with six Mo or W atoms. The calculated lattice constants are $a=4.75$ {\AA}, $b=6.06$ {\AA}, $c=5.22$ {\AA} for Mo$_{2}$C and $a=4.75$ {\AA}, $b=6.10$ {\AA}, $c=5.23$ {\AA} for W$_{2}$C, which agree well with the experimental values \cite{Ref_Mo2C_LAT,Ref_W2C_LAT}. In Fig. \ref{fig1}(b), the bulk BZ along with the high-symmetry $k$ points and the projected 2D BZ of the (001) surface are displayed.

\begin{figure}[t]
\includegraphics[angle=0,scale=0.23]{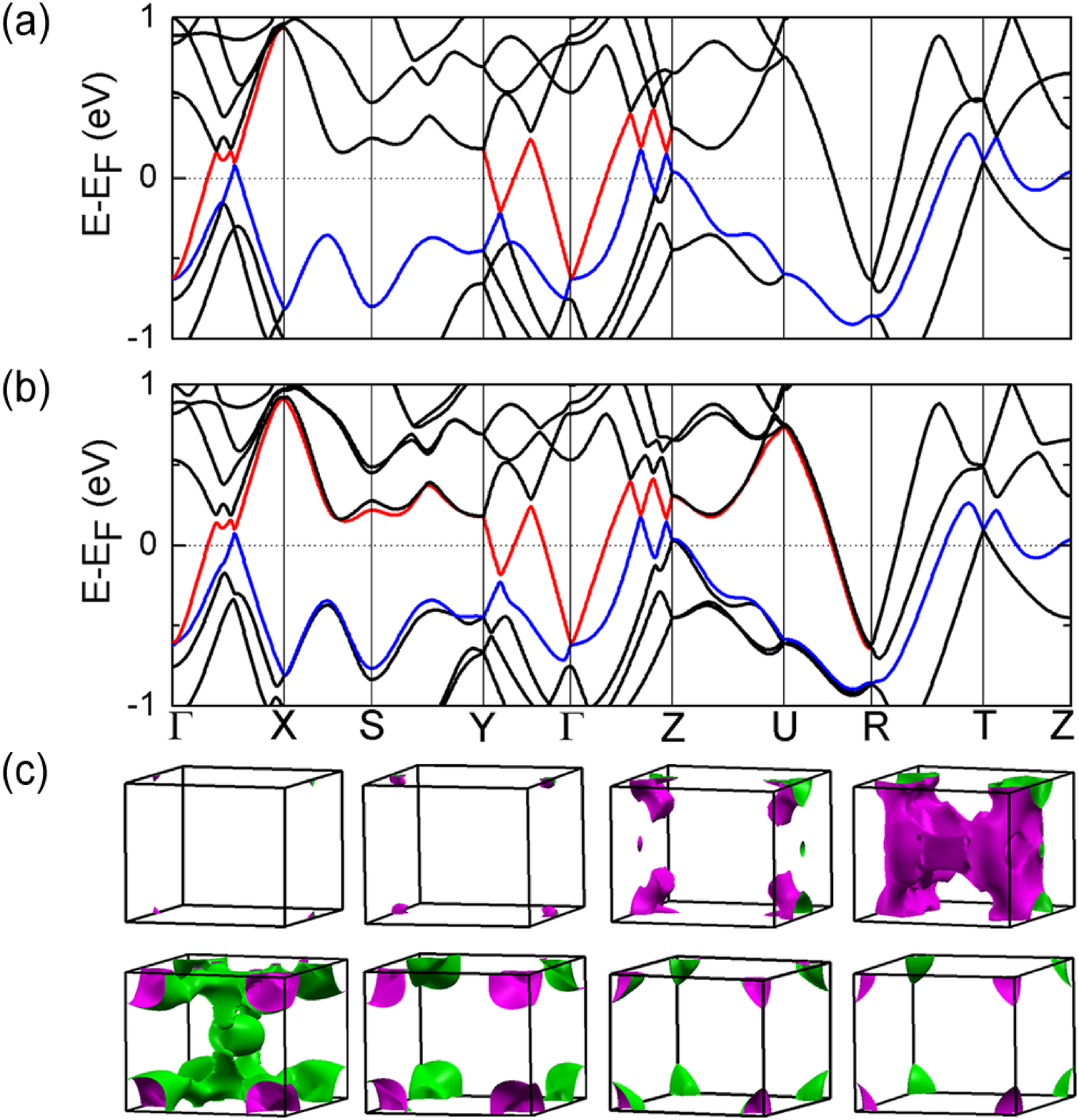}
  \caption{(Color online) Electronic band structures of Mo$_{2}$C calculated (a) without and (b) with the spin-orbit coupling (SOC) along the high-symmetry paths of BZ. (c) Fermi surface sheets of Mo$_{2}$C calculated without the SOC.  }
 \label{fig2}
\end{figure}

\begin{figure}[tbh]
\includegraphics[angle=0,scale=0.11]{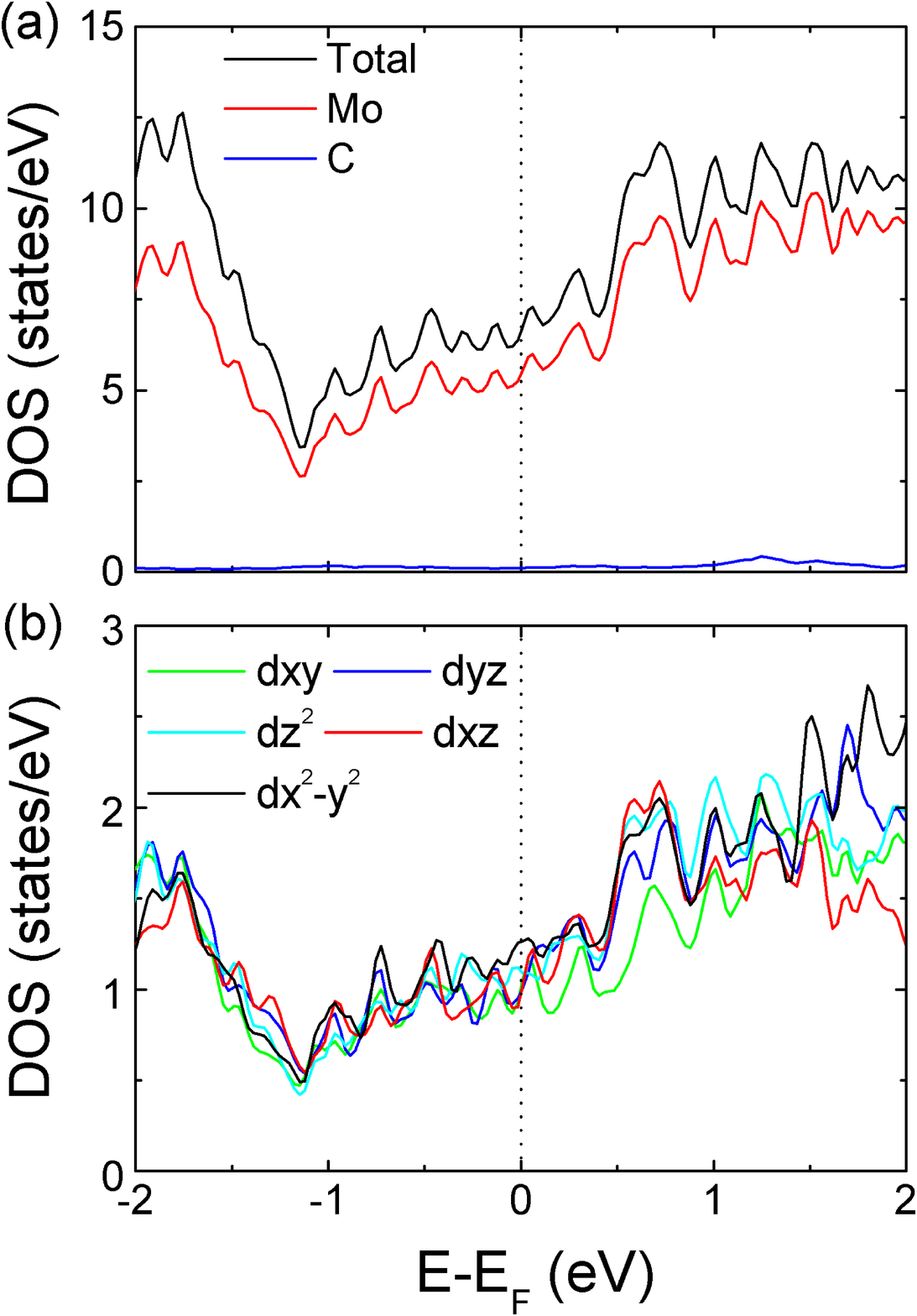}
  \caption{(Color online) (a) Total and local density of states for Mo$_{2}$C calculated without the SOC. (b) Partial density of states for five 4$d$ orbitals of Mo atom. The Fermi level sets to zero.}
  \label{fig3}
\end{figure}

\begin{figure}[tb]
\includegraphics[angle=0,scale=0.32]{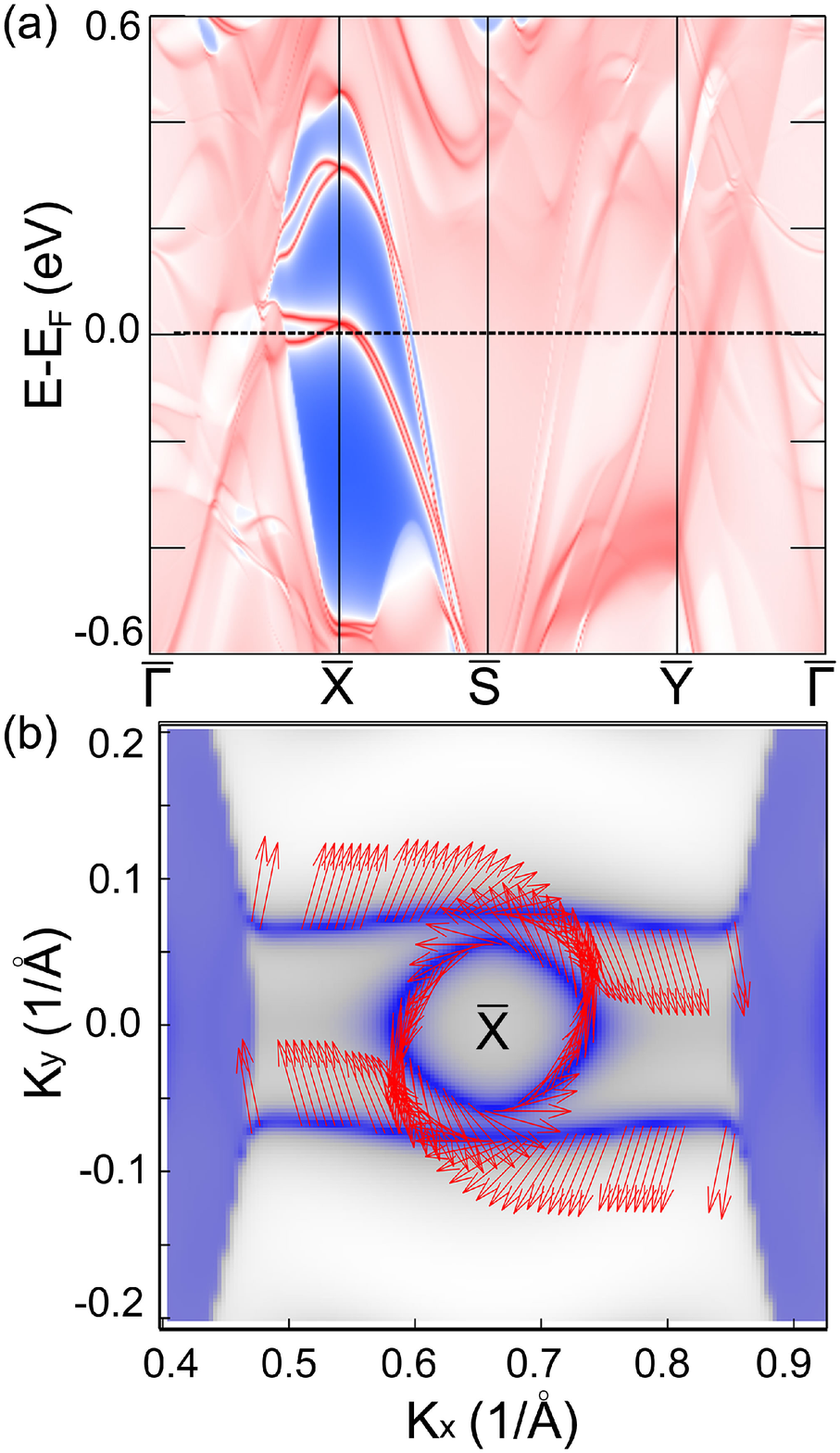}
  \caption {(Color online) (a) Band structure of the (001) surface of Mo$_{2}$C along the high-symmetry paths in the projected 2D BZ (Fig. \ref{fig1}) calculated with the SOC. (b) Surface states with spin textures around the $\overline{\mathrm{X}}$ point in the 2D BZ at a fixed energy of $E_\text{F}$. Here the arrows denote spins' directions.}
 \label{fig4}
\end{figure}

\begin{figure}[tb]
\includegraphics[angle=0,scale=0.30]{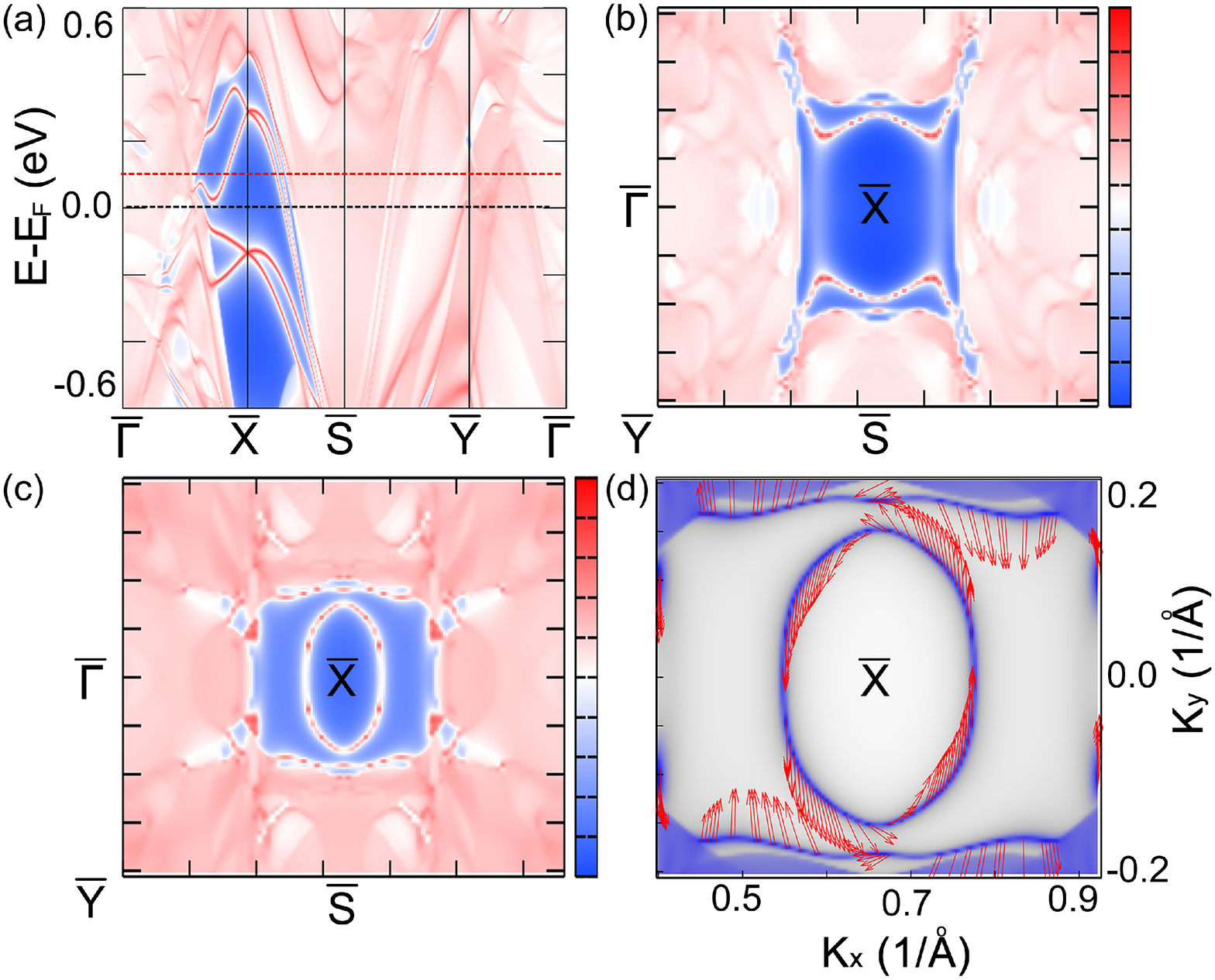}
  \caption{(Color online) (a) Band structure of the (001) surface of W$_{2}$C along the high-symmetry paths in the projected 2D BZ (Fig. \ref{fig1}) calculated with the SOC. (b) Surface states in the 2D BZ at a fixed energy of $E_\text{F}$. (c) Surface states and (d) spin textures around $\overline{\mathrm{X}}$ point in the 2D BZ at a fixed energy of $E_\text{F}$+0.1 eV. Here the arrows denote spins' directions.}
 \label{fig5}
\end{figure}

Figures \ref{fig2}(a) and \ref{fig2}(c) show the band structures along the high-symmetry paths of BZ and the Fermi surface of Mo$_{2}$C calculated without the spin-orbit coupling (SOC), respectively. From the band structure [Fig. \ref{fig2}(a)], we can see that there are several bands crossing the Fermi level with large dispersions, indicating the metallic behavior of Mo$_{2}$C. Moreover, the corresponding eight Fermi surface pockets shown in Fig. \ref{fig2}(c) demonstrate its three-dimensional (3D) character. According to the calculated density of states [Fig. \ref{fig3}(a)], Mo atoms have an essential contribution around the Fermi level in comparison with C atoms, which suggests that the superconductivity in Mo$_{2}$C can be attributed to Mo orbitals. The partial density of states shows that five 4$d$ orbitals of the Mo atom have similar weights around the Fermi level, which is due to the relatively weak crystal field effect in Mo$_{2}$C.

On the other hand, a careful examination of the band structure in Fig. \ref{fig2}(a) indicates that there are band inversions between the conduction bands and the valence bands, implying that Mo$_{2}$C may have a nontrivial topological property. Figure \ref{fig2}(b) shows the band structure of Mo$_{2}$C calculated with the SOC. When the SOC effect is included, the band crossings without the SOC [Fig. \ref{fig2}(a)] open with a continuous band gap appearing through the whole Brillouin zone [Fig. \ref{fig2}(b)], for which one opened gap is 11 meV at the $\Gamma$ point and two opened gaps are 13 meV and 14 meV around the Z point, respectively. This is because the highest rotationally symmetric operation of orthorhombic-phase Mo$_{2}$C is C$_{2}$, which cannot protect the band crossing points along the high-symmetry path\cite{Ref_Cd3As2}. As a result, a curved Fermi level can be defined between the highest valence band (blue color) and the lowest conduction band (red color), and we can further calculate the topological invariant for the occupied bands below the gap to check the topological property of Mo$_{2}$C. For a 3D system, the topological invariant is defined\cite{Ref_3DTI} as $Z_{2}$ ($\nu_{0};\nu_{1} \nu_{2} \nu_{3}$), where $\nu_{0}$ is a strong topological index and $(\nu_{1},\nu_{2},\nu_{3})$ are three weak topological indices\cite{Ref_3DTI_Is}. By using  the Wilson loop method\cite{Ref_Z2cal}, we obtain the topological invariant $Z_{2}$ as $(1; 000)$, which indicate that the orthorhombic-phase Mo$_{2}$C is a strong topological insulator defined on a curved Fermi surface (see Appendix A for details).

One of the most significant phenomena for topologically nontrivial materials is the existence of topological surface states. Figure \ref{fig4}(a) shows the band structure of the (001) surface of Mo$_{2}$C along the high-symmetry paths in the projected 2D BZ (Fig. \ref{fig1}) calculated with the SOC. The Dirac-type surface states located in the bulk band gap at the $\overline{\mathrm{X}}$ point can be clearly resolved, which can be detected by angle-resolved photoemission spectroscopy (ARPES) experiment. The surface states just pass through the Fermi level, giving rise to an ellipse surrounding the $\overline{\mathrm{X}}$ point. These surface states can hold spin-momentum locked spin textures as shown in Fig. \ref{fig4}(b), which are protected by the time-reversal symmetry \cite{Ref_TI}.

We have also studied the electronic structure of the orthorhombic-phase W$_{2}$C, which has the same crystal structure as Mo$_{2}$C. Due to the heavier atomic mass, the W atom has spin-orbit coupling stronger than that of the Mo atom. There are seven bands across the Fermi level (not shown) and the calculated topological invariant Z$_{2}$ of the orthorhombic-phase W$_{2}$C is (1;000), indicating that W$_{2}$C is also a strong topological insulator defined on a curved Fermi surface. Figure \ref{fig5}(a) shows the band structure of the (001) surface and the surface states of W$_{2}$C. Since there are surface states merging into the bulk states without crossing the Fermi level along the $\overline{\Gamma}$ - $\overline{\mathrm{X}}$ path, the surface states do not form a closed contour around the $\overline{\mathrm{X}}$ point as shown in Fig. \ref{fig5}(b). Nevertheless, when the Fermi level is lifted by 0.1 eV via electron doping, the surface states will pass through the Fermi level, giving rise to an ellipse surrounding the $\overline{\mathrm{X}}$ point [Fig. \ref{fig5}(c)]. Likewise, these surface states can hold spin-momentum locked spin textures as shown in Fig. \ref{fig5}(d). Since both Mo$_{2}$C and W$_{2}$C are strong topological insulators defined on curved Fermi surfaces, all of their surfaces exhibit topological surface states. As illustrations, we have also investigated the band structures of the (100) and (010) surfaces for Mo$_{2}$C and W$_{2}$C (see Fig. \ref{fig7} in Appendix B), where the topological surface states in bulk band gaps can be clearly resolved as well.

\section{Discussion and Summary}

The exploration for a single compound possessing  both superconductivity and nontrivial topological properties is important for the realization of topological superconductivity. According to Fu and Kane's proposal\cite{Ref_6}, once the bulks of Mo$_{2}$C and W$_{2}$C become superconducting below the critical temperatures, their spin-momentum locked surface states around the Fermi level will open a superconducting gap via the proximity effect. The spin-polarized surface electrons that can be regarded as spinless fermions will thus form the Cooper pairs. In this case, the resulting superconductivity on the surface can be viewed as an equivalent $p+ip$ superconductivity. When an appropriate external magnetic field is applied, Majorana zero modes will be localized in the vortex cores and the zero-bias peak can be observed in a scanning tunneling microscopy (STM) experiment. Our theoretical proposals on Mo$_{2}$C and W$_{2}$C need to be confirmed by further experimental studies.

In real materials, although the Dirac-type surface states are complicated, the key factor is the formation of spin-momentum locked spin textures as shown in Figs. \ref{fig4} and \ref{fig5}. To be a single-compound topological superconductor candidate, we propose that it meet the following criteria: (i) the nontrivial topological surface states locate in the projected bulk band gap and are not buried by bulk states; (ii) these surface states are close to the Fermi level and hold the helical spin texture around a time-reversal invariant point; (iii) the superconducting transition temperature of the compound should be as high as possible so that the Majorana zero modes can be separated from other Caroli-de Gennes-Matricon (CdGM) states in the superconducting gap\cite{Ref_Cdgm}. In previous studies, several kinds of single compounds have been predicted to be topological superconductors, such as FeTe$_{0.55}$Se$_{0.45}$ \cite{Ref_FeSeTecal1,Ref_FeSeTecal2}, LiFe$_{1-x}$Co$_{x}$As\cite{Ref_LiFeAs}, (Li$_{0.84}$Fe$_{0.16}$)OHFeSe\cite{Ref_LiFeOHFeSe}, CaKFe$_{4}$As$_{4}$\cite{Ref_CaKFe4As4}, TaSe$_{3}$\cite{Ref_TaSe3}, $A$15 system\cite{Ref_A15}, PdBi\cite{Ref_PdBi}, $\beta$-PdBi$_{2}$\cite{Ref_PdBi2,Ref_PdBi2-ex}, $\beta$-RhPb$_{2}$\cite{Ref_RhPb2}, YPb$_{3}$\cite{Ref_YD3}, Mg$_{2}$Pb\cite{Ref_Mg2Pb}, Au$_{2}$Pb\cite{Ref_Au2Pb}, LuPtBi\cite{Ref_LuPtBi}, PbTaSe$_{2}$\cite{Ref_PbTaSe2}, and PbTiSe$_{2}$\cite{Ref_PbTiSe2}, etc. In this work, we propose that Mo$_{2}$C and W$_{2}$C are also ideal topological superconductor candidates. First, their projected bulk states on the (001) surface have relatively large band gaps (about 1 eV) along the $\overline{\Gamma}$-$\overline{\mathrm{X}}$-$\overline{\mathrm{S}}$ path. As a result, the surface states can be easily detected and resolved in the ARPES experiment. Second, the surface states close to the Fermi level readily become superconducting via a proximity effect. The topological surface states of Mo$_{2}$C just cross the Fermi level, while those of W$_{2}$C would cross the Fermi level with slight electron doping of 0.8-1.4 electrons per formula unit to lift the Fermi level by 0.1-0.2 eV. Third, the superconducting transition temperature of Mo$_{2}$C is 7.3 K, higher than the liquid helium temperature and the $T_c$'s of many other topological superconductor candidates.

In summary, we have investigated the topological electronic properties and surface states of the orthorhombic-phase Mo$_{2}$C and W$_{2}$C superconductors by using first-principles electronic structure calculations. The calculated topological invariant Z$_{2}$ of (1;000) indicates that they are both strong topological insulators defined on curved Fermi levels. The nontrivial topological surface states are within the bulk band gaps and these Dirac-type surface states hold helical spin textures. Considering the fact that Mo$_{2}$C and W$_{2}$C had been reported to be superconductors in experiments, their bulk superconductivity may induce superconductivity in the nontrivial topological surface states via a proximity effect. Thus the orthorhombic-phase Mo$_{2}$C and W$_{2}$C provide an appropriate platform for investigating the interplay between superconductivity and topological properties as well as for manipulating Majorana zero modes in future experiments\cite{Ref_MFB}.

\section{APPENDIX A}

\begin{figure}[tb]
\includegraphics[angle=0,scale=0.32]{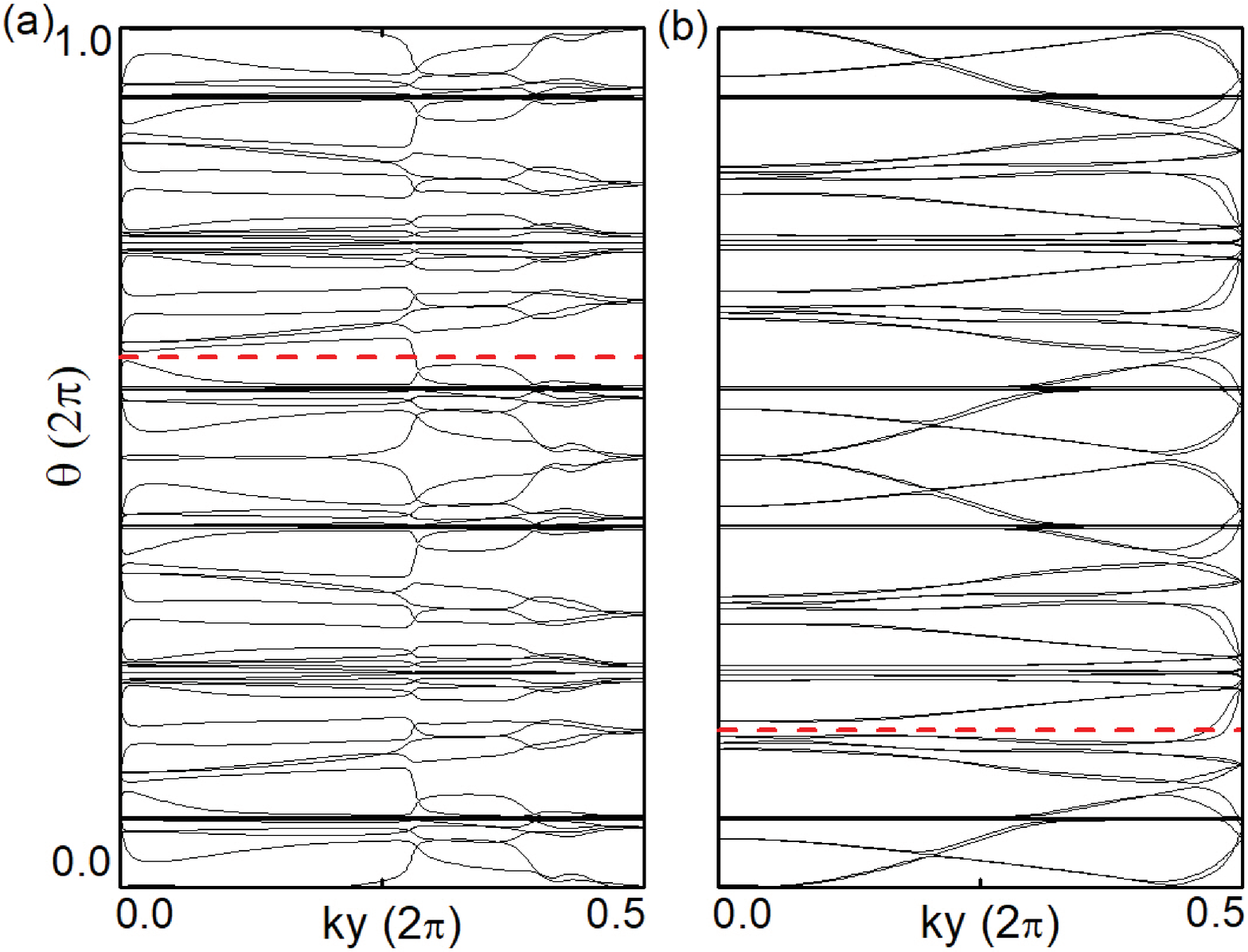}
\caption{(Color online) The evolution lines of Wannier centers of bulk Mo$_{2}$C in the (a) $k_x=0$ and (b) $k_x=\pi$ planes, respectively.}
 \label{fig6}
\end{figure}

We performed the calculations on the nontrivial topological invariants of Mo$_{2}$C in the presence of SOC~\cite{Ref_Z2cal}. Figures \ref{fig6}(a) and \ref{fig6}(b) show the evolution lines of Wannier centers in the $k_{x}$=0 and $k_{x}$=$\pi$ planes of the bulk BZ of Mo$_{2}$C, respectively. The evolution lines cross the reference line (red dashed line) an odd number of times
in the $k_{x}$=0 plane with the Z$_{2}$ index being 1 and an even number of times in the $k_{x}$=$\pi$ plane with Z$_{2}$ index being 0, leading to the strong topological index $\nu_{0}=1$. Therefore, Mo$_{2}$C is a 3D strong topological insulator. Likewise, W$_{2}$C is a 3D strong topological insulator. We have checked the topological properties by using a recently developed functional at the meta-GGA level, namely the strongly constrained and appropriately normed semilocal density Functional~\cite{scan}, which gives the consistent results.

\section{APPENDIX B}

\begin{figure}[tb]
\includegraphics[angle=0,scale=0.09]{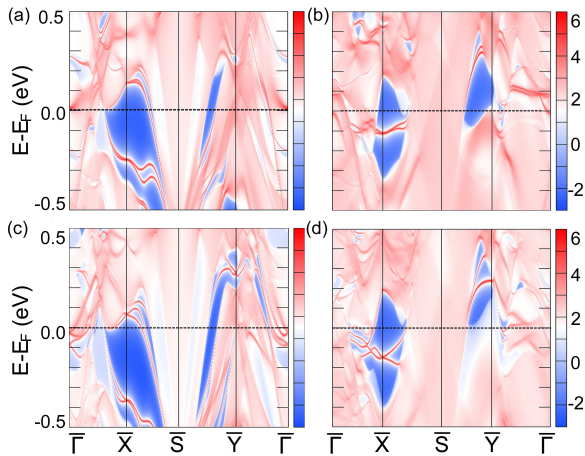}
 \caption{(Color online) Band structures of the (a) (010) and (b) (100) surfaces of Mo$_{2}$C, respectively. Band structures of the (c) (010) and (d) (100) surfaces of W$_{2}$C, respectively.}
 \label{fig7}
\end{figure}

Here we show the surface states of Mo$_{2}$C and W$_{2}$C on another two surfaces. Figures \ref{fig7}(a) and \ref{fig7}(b) respectively show the band structures of the (010) and (100) surfaces of Mo$_{2}$C. Figures \ref{fig7}(c) and \ref{fig7}(d) respectively show those of W$_{2}$C. The surface states in the projected bulk band gaps around the $X$ point of the 2D BZ can be clearly resolved.

\begin{acknowledgments}
This work was supported by the National Key R\&D Program of China (Grants No. 2017YFA0302903 and No. 2019YFA0308603), the National Natural Science Foundation of China (Grants No. 11774422 and No. 11774424), the CAS Interdisciplinary Innovation Team, the Fundamental Research Funds for the Central Universities, and the Research Funds of Renmin University of China (Grant No. 19XNLG13).  Computational resources were provided by the Physical Laboratory of High Performance Computing at Renmin University of China.
\end{acknowledgments}


\begin{thebibliography}{}

\bibitem{Ref_1} X.-L. Qi and S.-C. Zhang, Rev. Mod. Phys. {\bf 83}, 1057 (2011).
\bibitem{Ref_2} M. Sato and Y. Ando, Rep. Prog. Phys. {\bf 80}, 076501 (2017).
\bibitem{Ref_3}  J. Alicea, Rep. Prog. Phys. {\bf 75}, 076501 (2012).
\bibitem{Ref_4}  C. Nayak, S. H. Simon, A. Stern, M. Freedman, and S. Das Sarma, Rev. Mod. Phys. {\bf 80}, 1083 (2008).
\bibitem{Ref_5}  A. P. Mackenzie and Y. Maeno, Rev. Mod. Phys. {\bf 75}, 657 (2003).
\bibitem{Ref_6}  L. Fu and C. L. Kane, Phys. Rev. Lett. {\bf 100}, 096407 (2008).
\bibitem{Ref_7Nanowire} V. Mourik, K. Zuo, S. M. Frolov, S. Plissard, E. P.
Bakkers, and L. P. Kouwenhoven, Science {\bf 336}, 1003 (2012).
\bibitem{Ref_Bi2Se3} M.-X. Wang, C. Liu, J.-P. Xu, F. Yang, L. Miao, M.-Y. Yao, C. L. Gao, C. Shen, X. Ma, X. Chen, Z.-A. Xu, Y. Liu, S.-C. Zhang, D. Qian, J.-F. Jia, and Q.-K. Xue, Science {\bf 336}, 52 (2012).
\bibitem{Ref_8Featom}  S. Nadj-Perge, I. K. Drozdov, J. Li, H. Chen, S. Jeon, J. Seo, A. H. MacDonald, B. A. Bernevig, and A. Yazdani, Science {\bf 346}, 602 (2014).
\bibitem{Ref_9Bi2Se3/NbSe2}J.-P. Xu, M.-X. Wang, Z.-L. Liu, J.-F. Ge, X. Yang, C. Liu, Z.
-A. Xu, D. Guan, C.-L. Gao, D. Qian, Y. Liu, Q.-H. Wang, F.-C. Zhang, Q.-K. Xue, and J.-F. Jia, Phys. Rev. Lett. {\bf 114}, 017001 (2015).
\bibitem{Ref_exFeSeTe1} J.-X. Yin, Z. Wu, J.-H. Wang, Z.-Y. Ye, J. Gong, X.-Y. Hou, L. Shan, A. Li, X.-J. Liang, X.-X. Wu, J. Li, C.-S. Ting, Z.-Q. Wang, J.-P. Hu, P.-H. Hor, H. Ding, and S. H. Pan, Nat. Phys, {\bf 11}, 543 (2015).
\bibitem{Ref_exFeSeTe2} D.-F. Wang, L.-Y. Kong, P. Fan, H. Chen, S.-Y. Zhu, W.-Y. Liu, L. Cao, Y.-J. Sun, S.-X. Du, J. Schneeloch, R.-D. Zhong, G.-D. Gu, L. Fu, H. Ding, and H.-J. Gao, Science {\bf362}, 333 (2018).
\bibitem{Ref_exFeSeTe3} P. Zhang, K. Yaji, T. Hashimoto, Y. Ota, T. Kondo, K. Okazaki, Z. Wang, J. Wen, G. D. Gu, H. Ding, and S. Shin, Science {\bf 360}, 182 (2018).
\bibitem{Ref_LiFeOHFeSe} Q. Liu, C. Chen, T. Zhang, R. Peng, Y.-J. Yan, C.-H.-P. Wen, X. Lou, Y.-L. Huang, J.-P. Tian, X.-L. Dong, G.-W. Wang, W.-C. Bao, Q.-H. Wang, Z.-P. Yin, Z.-X. Zhao, and D.-L. Feng, Phys. Rev. X {\bf 8}, 041056 (2018).
\bibitem{Ref_10} L. A. Wray, S.-Y. Xu, Y. Xia, Y. S. Hor, D. Qian, A. V. Fedorov, H. Lin, A. Bansil, R. J. Cava, and M. Z. Hasan, Nat. Phys. {\bf 6}, 855 (2010).
\bibitem{Ref_11_MFvortice} P. Hosur, P. Ghaemi, R. S. K. Mong, and A. Vishwanath, Phys. Rev. Lett. {\bf 107}, 097001 (2011).
\bibitem{Ref_CuxBi2Se3} S. Sasaki, M. Kriener, K. Segawa, K. Yada, Y. Tanaka, M. Sato, and Y. Ando, Phys. Rev. Lett. {\bf 107}, 217001 (2011).
\bibitem{Ref_NbxBi2Se3}T. Asaba, B. J. Lawson, C. Tinsman, L. Chen, P. Corbae, G. Li, Y. Qiu, Y. S. Hor, L. Fu, and L. Li, Phys. Rev. X {\bf 7}, 011009 (2017).
\bibitem{Ref_12syn} M. Kriener, K. Segawa, Z. Ren, S. Sasaki, S. Wada, S. Kuwabata, and Y. Ando, Phys. Rev. B {\bf 84}, 054513 (2011).
\bibitem{Ref_Sb2Te3_pres} J. Zhu, J.-L. Zhang, P.-P. Kong, S.-J. Zhang, X.-H. Yu, J.-L. Zhu, Q.-Q. Liu, X. Li, R.-C. Yu, R. Ahuja, W.-G. Yang, G.-Y. Shen, H.-K. Mao, H.-M. Weng, X. Dai, Z. Fang, Y.-S. Zhao, and C.-Q. Jin, Sci. Rep. {\bf 3}, 2016 (2013).
\bibitem{Ref_Cd3As2_SC} H. Wang, H.-C. Wang, H.-W. Liu, H. Lu, W.-H. Yang, S. Jia, X.-J. Liu, X.-C. Xie, J. Wei, and J. Wang, Nat. Mater. {\bf 15}, 38 (2016).
\bibitem{Ref_FeSeTe_film} X.-X. Wu, S.-S. Qin, Y. Liang, H. Fan, and J.-P. Hu, Phys. Rev. B {\bf 93}, 115129 (2016).
\bibitem{Ref_FeSeTecal1} G. Xu, B. Lian, P.-Z. Tang, X.-L. Qi, and S.-C. Zhang, Phys. Rev. Lett. {\bf 117}, 047001 (2016).
\bibitem{Ref_FeSeTecal2} Z.-J. Wang, P. Zhang, G. Xu, L. K. Zeng, H. Miao, X.-Y. Xu, T. Qian, H.-M. Weng, P. Richard, A. V. Fedorov, H. Ding, X. Dai, and Z. Fang, Phys. Rev. B  {\bf 92}, 115119 (2015).
\bibitem{Ref_LiFeAs} P. Zhang, Z.-J. Wang, X.-X. Wu, K. Yaji, Y. Ishida, Y. Kohama, G.-Y. Dai, Y. Sun, C. Bareille, K. Kuroda, T. Kondo, K. Okazaki, K. Kindo, X.-C. Wang, C.-Q. Jin, J.-P. Hu, R. Thomale, and K. Sumida, Nat. Phys. {\bf 15}, 41 (2019).
\bibitem{Ref_Tc_Mo} N. Morton, B. W. James, G. H. Wostenholm, D. G. Pomfret, and M. R. Davies, J. Less Common Met. {\bf 25}, 97 (1971).
\bibitem{Ref_Tc_W} N. Morton, B. W. James, G. H. Wostenholm, D. G. Pomfret, and M. R. Davies, J. Less Common Met.  {\bf 29}, 423 (1972).
\bibitem{Ref_Dft} W. Kohn and L. J. Sham, Phys. Rev. {\bf 140}, A1133 (1965).
\bibitem{Ref_paw} P. E. Bl$\ddot{\mathrm{o}}$chl, Phys. Rev. B {\bf 50}, 17953 (1994).
\bibitem{Ref_vasp1} G. Kresse and J. Hafner, Phys. Rev. B {\bf 47}, 558 (1993).
\bibitem{Ref_vasp2} G. Kresse and J. Furthm\"uller, Phys. Rev. B {\bf 54}, 11169 (1996).
\bibitem{Ref_gga} J. P. Perdew, K. Burke, and M. Ernzerhof, Phys. Rev. Lett. {\bf 77}, 3865 (1996).
\bibitem{Ref_wantool} Q. Wu, S. Zhang, H.-F. Song, M. Troyer, and A. A. Soluyanov, Comput. Phys. Commun. {\bf 224}, 405 (2018).
\bibitem{Ref_wanfun1}  N. Marzari, A. A. Mostofi, J. R. Yates, I. Souza, and D. Vanderbilt, Rev. Mod. Phys. {\bf 84} 1419 (2012).
\bibitem{Ref_wanfun2} A. A. Mostofi, J. R. Yates, G. Pizzi, Y.-S. Lee, I. Souza, D. Vanderbilt, and N. Marzari, Comput. Phys. Commun. {\bf185} 2309 (2014).
\bibitem{Ref_Mo2C_LAT}T. Epicier, J. Dubois, C. Esnouf, G. Fantozzi, and P. Convert, Acta Metall. {\bf 36}, 1903 (1988).
\bibitem{Ref_W2C_LAT}K. Yvon, H. Nowotny, and F. Benesovsky, Monatsch. Chem. {\bf 99}, 726 (1968).
\bibitem{Ref_Cd3As2} Z. Wang, H. Weng, Q. Wu, X. Dai, and Z. Fang, Phys. Rev. B {\bf 88}, 125427 (2013).
\bibitem{Ref_3DTI} L. Fu, C. L. Kane, and E. J. Mele, Phys. Rev. Lett. {\bf 98}, 106803 (2007).
\bibitem{Ref_3DTI_Is}  L. Fu and C. L. Kane, Phys. Rev. B {\bf 76}, 045302 (2007).
\bibitem{Ref_Z2cal} R. Yu, X.-L. Qi, A. Bernevig, Z. Fang, and X. Dai, Phys. Rev. B {\bf 84}, 075119 (2011).
\bibitem{Ref_TI}  M. Z. Hasan and C. L. Kane, Rev. Mod. Phys. {\bf 82}, 3054 (2010).
\bibitem{Ref_Cdgm} M.-Y. Chen, X.-Y. Chen, H. Yang, Z.-Y. Du, X.-Y. Zhu, E.-Y. Wang, and H.-H. Wen, Nat. Commun. {\bf 9}, 970 (2018).
\bibitem{Ref_CaKFe4As4} W. Liu, L. Cao, S. Zhu, L. Kong, G. Wang, M. Papaj, P. Zhang, Y. Liu, H. Chen, G. Li, F. Yang, T. Kondo, S. Du, G. Cao, Shik. Shin, L. Fu, Z. Yin, H.-J. Gao, and H. Ding, arXiv: 1907, 00904 (2019).
\bibitem{Ref_TaSe3} S. Nie, L. Xing, R. Jin, W. Xie, Z. Wang, and F. B. Prinz, Phys. Rev. B {\bf 98}, 125143 (2018).
\bibitem{Ref_A15} M. Kim, C.-Z. Wang, and K.-M. Ho, Phys. Rev. B {\bf 99}, 224506 (2019).
\bibitem{Ref_PdBi} M. Neupane, N. Alidoust, M. M. Hosen, J.-X. Zhu, K. Dimitri, S.-Y. Xu, N. Dhakal, R. Sankar, I. Belopolski, D. S. Sanchez, T.-R. Chang, H.-T. Jeng, K. Miyamoto, T. Okuda, H. Lin, A. Bansil, D. Kaczorowski, F. Chou, M. Z. Hasan, and T. Durakiewicz, Nat. Commun. {\bf 7}, 13315 (2016).
\bibitem{Ref_PdBi2}  M. Sakano, K. Okawa, M. Kanou, H. Sanjo, T. Okuda, T. Sasagawa, and K. Ishizaka, Nat. Commun. {\bf 6}, 8595 (2015).
\bibitem{Ref_PdBi2-ex}  Y.-F. Lv, W.-L. Wang, Y.-M. Zhang, H. Ding, W. Li, L. Wang, K. He, C.-L. Song, X.-C. Ma, and Q.-K. Xue, Sci. Bull. {\bf 62}, 852 (2017).
\bibitem{Ref_RhPb2}J.-F. Zhang, P.-J .Guo, M. Gao, K. Liu, and Z.-Y. Lu, Phys. Rev. B {\bf 99}, 045110 (2019).
\bibitem{Ref_YD3} X.-H. Tu, P.-F. Liu, and B.-T. Wang, Phys. Rev. Mater. {\bf 3}, 054202 (2019).
\bibitem{Ref_Mg2Pb} G. Bian, T.-R. Chang, A. Huang, Y. Li, H.-T. Jeng, D. J. Singh, R. J. Cava, and W. Xie, Phys. Rev. Mater. {\bf 1}, 021201 (2017).
\bibitem{Ref_Au2Pb} Y. Xing, H. Wang, C.-K. Li, X. Zhang, J. Liu, Y. Zhang, J. Luo, Z. Wang, Y. Wang, L. Ling, M. Tian, S. Jia, J. Feng, X.-J. Liu, J. Wei, and J. Wang, npj Quant. Mater. {\bf 1}, 16005 (2016).
\bibitem{Ref_LuPtBi} F. Tafti, T. Fujii, A. Juneau-Fecteau, S. R. de Cotret, N. Doiron-Leyraud, A. Asamitsu, and L. Taillefer, Phys. Rev. B {\bf 87},  184504 (2013).
\bibitem{Ref_PbTaSe2}  T.-R. Chang, P.-J. Chen, G. Bian, S.-M. Huang, H. Zheng, T. Neupert, R. Sankar, S.-Y. Xu, I. Belopolski, G. Chang, B. K. Wang, F. Chou, A. Bansil, H.-T. Jeng, H. Lin, and M. Z. Hasan, Phys. Rev. B {\bf 93}, 245130 (2016).
\bibitem{Ref_PbTiSe2} P.-J. Chen, T.-R. Chang, and H.-T. Jeng, Phys. Rev. B {\bf 94}, 165148 (2016).
\bibitem{Ref_MFB}S. Das Sarma, M. Freedman, and C. Nayak, Quantum. Inf. {\bf 1}, 15001 (2015).
\bibitem{scan}J. W. Sun, A. Ruzsinszky, and J. P. Perdew, Phys. Rev. Lett. {\bf 115}, 036402 (2015).

\end{thebibliography}
\end{document}